\documentclass[aps,pra,reprint,superscriptaddress,showkeys]{revtex4-1}

\usepackage{subcaption}
\usepackage{graphicx}
\usepackage{float}
\usepackage{romannum}
\usepackage{csquotes}
\usepackage{amsmath,amsthm,mathtools}
\usepackage[dvipsnames]{xcolor}
\usepackage{ulem}
\usepackage{caption}
\renewcommand{\eqref}[1]{Eq.~\ref{#1}}

\begin{document}

\author{Jakub Szlachetka}

\thanks{These authors contributed equally to this work}

\email{Corresponding author: jakub.szlachetka@doktorant.umk.pl}

\affiliation{Institute of Physics, Faculty of Physics, Astronomy and Informatics \\ Nicolaus Copernicus University, Grudziadzka 5, 87--100 Torun, Poland}

\author{Kaushik Joarder}
\thanks{These authors contributed equally to this work}

\affiliation{Institute of Physics, Faculty of Physics, Astronomy and Informatics \\ Nicolaus Copernicus University, Grudziadzka 5, 87--100 Torun, Poland}

\author{Piotr Kolenderski }
\affiliation{Institute of Physics, Faculty of Physics, Astronomy and Informatics \\ Nicolaus Copernicus University, Grudziadzka 5, 87--100 Torun, Poland}

\title{Ultrabright Sagnac-type source of non-degenerate polarization-entangled photon pairs using only off-the-shelf optical components }

\begin{abstract}
We develop a Sagnac-type source of ultrabright, non-degenerate, polarization-entangled photon pair that is highly stable and compact simultaneously. We use a $20~\text{mm}$ long PPLN bulk crystal which, upon pumping continuously with $532~\text{nm}$ wavelength, produces polarization-entangled photon-pairs of $785~\text{nm}$ signal and $1651~\text{nm}$ idler wavelengths via the type-0 spontaneous parametric down-conversion (SPDC) process. All optical components used in the setup are off-the-shelf components, readily available commercially; hence, no custom-designed or costly multi-wavelength polarization optics are required. At the same time, long-term phase stability is achieved without any additional active stabilization; due to the geometry of our Sagnac-type design. We also report one of the highest brightness of non-degenerate polarization-entangled photon pairs available in the literature. Even with a very low pump power of $0.034~\text{mW}$, we detect a coincidence rate of $(6.96\pm 0.03)\times10^{4}$ entangled pairs/sec/mW (averaged over three polarization-basis measurements: H/V, D/A, and R/L basis). The source's brightness is calculated to be $(6.17\pm 0.04)\times10^{6}$ entangled pairs/sec/mW for the signal line width of $0.4~\text{nm}$. From the raw coincidence counts (without any background coincidence correction), the fidelity of the entangled state is measured to be $(96.72\pm 0.01)\%$ with a concurrence of $(94.68\pm 0.20)\%$. Bell-CHSH inequality violation is reported as $S=2.71\pm 0.06$.
\end{abstract}

\maketitle

\section{Introduction}
\label{sec:intro}  
Quantum entanglement \cite{horodecki2009quantum} is a purely quantum mechanical phenomenon where two or more entangled systems show non-classical correlations which can be exploited to perform various tasks that are impossible in the realm of classical mechanics, for example, quantum teleportation \cite{PhysRevLett.70.1895}, superdense coding \cite{PhysRevLett.69.2881}, quantum key distribution (QKD)\cite{PhysRevLett.67.661, PhysRevLett.68.557}, etc. For the experimental demonstration of such tasks, highly entangled photon-pair sources are being developed using various technologies, as well as using different degrees of freedom for entangling the photon pairs, such as polarization \cite{steinlechner2013phase}, temporal \cite{marcikic2004distribution}, spatial degree of freedom \cite{walborn2010spatial}, etc. Among the different types of entangled-photon generation processes spontaneous parametric down-conversion (SPDC) has been widely used \cite{doi:10.1063/5.0023103}, where the polarizations of the two photons are most commonly entangled with each other.
\par
In the context of the current demand for technical and research purposes, a non-degenerate polarization-entangled photon source with high performance has become a necessary tool. In this source type, entangled photon pairs are generated in two vastly separated central wavelengths. Typically the signal wavelength is in the range of $750~\text{nm}$ to $850~\text{nm}$, and the idler wavelength is in the range of $1500~\text{nm}$ to $1700~\text{nm}$. This type of source is also called a \enquote{three-color} source, as all three operating wavelengths (pump, signal, and idler wavelengths) are significantly different from each other. The signal wavelength range is suitable for the highest detection efficiency in typical single-photon avalanche detectors (SPADs) that are readily available commercially. Also, it belongs to a lower absorption window in our atmosphere. On the other hand, the idler wavelength range is well within the optimal transmission window of commercial-grade optical fibers. Currently available superconducting detectors (SNSPDs) also provide higher quantum efficiency for this idler-wavelength regime. The advantage of using non-degenerate entangled-photon pairs is significant in the long-distance (satellite-to-earth) quantum communication channel (QKD), where the signal wavelength is most suited for satellite-to-ground-station communication \cite{yin2017satellite}, and the idler wavelength is efficient for terrestrial optical-fiber-based communication networks (intra-city and inter-city links) \cite{valivarthi2016quantum}. Another area of application is related to quantum memory, where the narrow-band signal photons are suitable to store information in a quantum memory (trapped Rubidium (Rb) atoms in a cavity \cite{PhysRevA.71.033805}, Nd:YSO crystal \cite{Clausen_2014}), and the idler photons are ideal candidates for entanglement-distribution between long-distant nodes connected with optical fibers.
\par
A good source of entangled photons should ideally possess the four general properties; 1) high performance, 2) low resource requirement, 3) stability and 4) adaptability. High performance can be quantified by measuring various parameters like entangled-pair-generation rate or brightness of the source, detected coincidence rate, the fidelity of entanglement, concurrence, purity, polarization-visibility, etc. Low resource requirement implies a low cost of production as well as less weight and power consumption. For this criteria, a compact design of the source setup with a minimal number of optical components is desirable. Custom-designed optics requiring a cutting-edge technical solution is costly; hence all the optical components are desired to be off-the-shelf components easily accessible in a standard optics laboratory. Recent impetus on satellite-based communication also encourages the design of the miniaturized version of the source compatible with a nano-satellite (CubeSat), which puts more constraints on the weight and power consumption. The source also should be stable for a longer duration, meaning that the performance should not degrade significantly over time. External factors like temperature, humidity, and mechanical stress, as well as internal parameters like robustness and type of interferometer, may affect the stability. To counter this, an active stabilization mechanism can be implemented within the source setup. However, this introduces additional components and complexity in the design. Hence, setup design with in-built stability (for example, Sagnac interferometric setup \cite{PhysRevA.73.012316,fedrizzi2007wavelength}) is highly desirable. The source must be adaptable for various purposes with minimal modifications. For example, the source should perform in different pumping conditions (both continuous and pulsed pumping), for various non-linear mediums (bulk crystals, fibers, waveguides, etc.), and for all type-0, \Romannum{1}, and \Romannum{2} pair-generation processes. Frequencies (or wavelength) of the generated entangled pairs should be tunable for a broad range of values.
\par
There have been a few experimental demonstrations of non-degenerate polarization-entangled photon sources. These sources can be broadly categorized into three general types, namely, 1) single crystal-single pass, 2) double crystal-single pass, and 3) single crystal-double pass. In the first category \cite{deChatellus:06}, the pump beam of one polarization direction (extraordinary or $e$) is passed through a single non-linear medium, while the specially manufactured crystal allows producing two simultaneous parametric down-conversion processes ($\left|e\right\rangle_{p} \rightarrow \left|e\right\rangle_{s}\left|e\right\rangle_{i}$ and $\left|e\right\rangle_{p} \rightarrow \left|o\right\rangle_{s}\left|o\right\rangle_{i}$) at the same time. A coherent combination of the two processes generates the entangled state $( \left|e\right\rangle_{s}\left|e\right\rangle_{i}+ \left|o\right\rangle_{s}\left|o\right\rangle_{i})/\sqrt{2}$. Although this source is simpler, compact, and less resource intensive, it requires active phase stabilization and specially designed crystal which is costly and not readily available. Also, the non-collinear pair production reduces the coupling efficiency resulting in lower performance. The second type of source uses two separate but nearly-identical crystals to produce $\left|i\right\rangle_{s}\left|j\right\rangle_{i}$ and $\left|i^{\perp}\right\rangle_{s}\left|j^{\perp}\right\rangle_{i}$ states in a coherent way. Here, $\left|i^{\perp}\right\rangle$ represents the orthogonal polarization state to $\left|i\right\rangle$, and the state $\left|j\right\rangle$ is either identical or orthogonal to the polarization state $\left|i\right\rangle$ ($\left|j\right\rangle=\left|i\right\rangle$ or $\left|j\right\rangle=\left|i^{\perp}\right\rangle$). These two crystals can be put side by side while the second crystal is rotated $90^{\circ}$ with respect to the first crystal \cite{Pelton:04,Hubel:07} or the two crystals can be put in parallel \cite{Clausen_2014}. The main issue with this type of source is to manufacture two identical crystals such that the generated photon pairs from both crystals are indistinguishable in all degrees of freedom (spatial, temporal, spectral). Also, the use of two crystals increases the cost of production. In addition, for the side-by-side configuration, the crystal's length is limited to maintain temporal overlap of the generated photon wave packets, reducing the source's brightness. On the other hand, in the parallel configuration, a Mach-Zehnder (MZ) type interferometer is required for a coherent combination of the two output states from the two crystals, introducing additional complexity of active phase stabilization for achieving long-term phase stability. The third type of source requires only one non-linear crystal, whereas the pump beam excites the crystal twice to produce the coherent states $\left|i\right\rangle_{s}\left|j\right\rangle_{i}$ and $\left|i^{\perp}\right\rangle_{s}\left|j^{\perp}\right\rangle_{i}$. Two parallel pump beams can be passed simultaneously through the crystal \cite{Fiorentino:08}, or two pump beams can hit the crystal from opposite sides along the same line \cite{PhysRevA.71.033805, Sauge:08, Lee_2021, Hentschel:09}. In the first category \cite{Fiorentino:08}, a beam-displacer (BD) based MZ interferometer has been used for the coherent combination of the two states, which is very compact in design as well as highly stable as compared to the conventional MZ interferometer. However, one drawback is the necessary use of the custom-designed miniaturized waveplates in one of the arms, as the two parallel pump beams must be very closely separated to pass through a single crystal with limited width. In the second category, conventional MZ-type interferometers are used \cite{PhysRevA.71.033805, Sauge:08} that requires active phase stabilization, or a Sagnac-type interferometer is used \cite{ Lee_2021, Hentschel:09} which does not require any additional stabilization mechanism due to its geometry providing in-built phase compensation.
\par
To our understanding, a Sagnac-type setup \cite{ Lee_2021, Hentschel:09} is the most promising way to generate non-degenerate entangled photons. However, a few issues still need to be addressed to achieve all four properties of a good source, as discussed earlier. In a Sagnac design of the source, one half-waveplate must be placed in one arm of the interferometer, which converts horizontal polarization to vertical (and vice versa). If one places this waveplate inside the Sagnac loop, then the half-waveplate should work for all three operating wavelengths (pump, signal, and idler wavelengths), which can only be custom-designed and with limited performance as per the currently available technology. One may also put the waveplate outside the Sagnac loop, which then necessitates the use of a miniaturized half-waveplate. The authors in ref. \cite{Lee_2021} bypassed this issue altogether by using a fiber (PMF) medium instead of a $\chi^{2}$ non-linear crystal. One end of the fiber was rotated $90^{\circ}$ with respect to the other end, which enables the use of the half-waveplate completely unnecessary. However, the use of the four-wave-mixing process in PMF instead of the SPDC process in a non-linear crystal has its drawbacks, for example, Raman scattering and high thermal sensitivity of the PMF. Also, the PMF's high loss and low coupling efficiency limit the source's brightness.
\par
In our source design, we use a periodically poled lithium-niobate (PPLN) crystal which produces entangled photon pairs from the SPDC process, while a similar Sagnac-type configuration \cite{ Lee_2021} is implemented. To counter the half-waveplate issue, we prepare two optical periscopes inside the Sagnac design, which act as half-waveplates. The periscope is constructed from two protected silver mirrors that can rotate the input polarization by $90^{\circ}$; for a broad range of wavelengths ($450~\text{nm}$ to $20~\mu \text{m}$). The usage of periscopes instead of half-waveplates inside the Sagnac setup was introduced earlier in ref.\cite{Hentschel:09}. However, their Sagnac design required the usage of a custom-designed Glan-Thompson polarizer for compensating angular dispersion. In contrast, our entangled-photon source uses all off-the-shelf components readily available in an optical laboratory. No custom-designed or costly optical components are used. At the same time, our setup design provides long-term phase stability without any additional active stabilization just by exploiting the inherent stability of the Sagnac design. We use PPLN crystal for the SPDC-based photon-pair generation, which can achieve higher brightness at lower pump power, as compared to other conventional non-linear crystals (such as PPKTP, BBO), due to its higher $\chi^{2}$ non-linearity coefficient. PPLN allows tuning the wavelengths of the generated photon pairs for a broad range; by only controlling the temperature of the crystal. Also, the spectral bandwidth of the photons is smaller (order of hundreds of GHz) due to the non-degenerate down-conversion process \cite{steinlechner2012high}, which can be modified to be suited for cascading with atomic systems.

\section{Design considerations}
\label{sec:design_considerations}
\begin{figure*}
\centering
\includegraphics[width=0.8\textwidth]{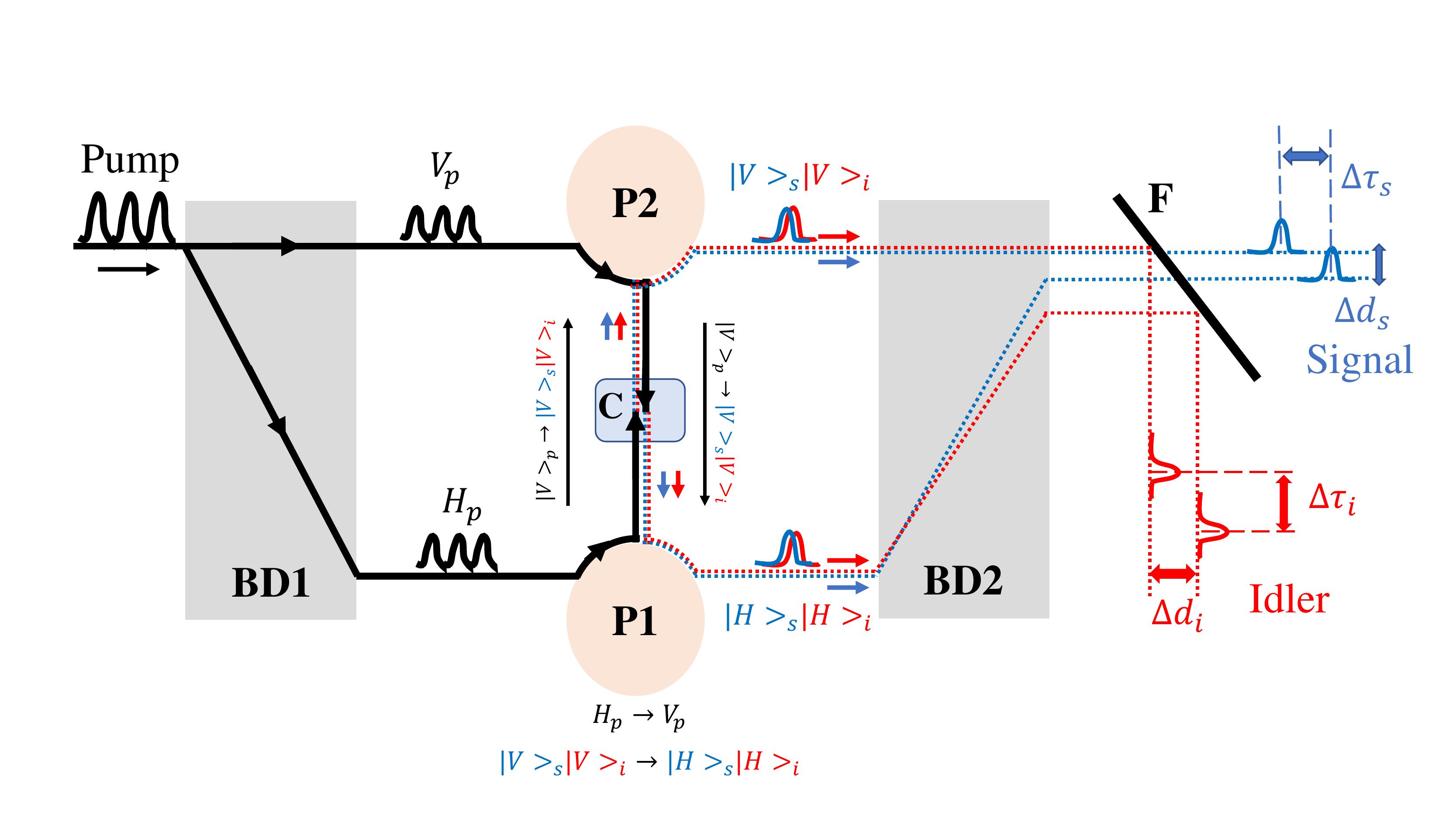}
\caption[example]
{ \label{fig:Sagnac_Design}
Conceptual design of the source. BD1, BD2: beam displacers; P1: periscope 1, which simultaneously converts the horizontally polarized pump beam to vertically polarized ($H_{p}\rightarrow V_{p}$) and vertically polarized signal and idler photon pairs to horizontally polarized ($\left|V\right\rangle_{s}\left|V\right\rangle_{i}\rightarrow\left|H\right\rangle_{s}\left|H\right\rangle_{i}$); P2: periscope 2 which maintains the input polarization of both the pump beam and the single-photon pairs; C: PPLN crystal for type-0 SPDC process ($\left|V\right\rangle_{p}\rightarrow\left|V\right\rangle_{s}\left|V\right\rangle_{i}$) in bidirectional continuous pumping condition; F: filter for separating the signal photons from the idler photons. $\Delta d_{s(i)}$ is the path separation, and $\Delta\tau_{s(i)}$ is the time delay between the two signal (idler) photon wave packets coming from bidirectional pumping.}
\end{figure*}

The basic design of our source is represented in Figure \ref{fig:Sagnac_Design}, which is conceptually similar to a beam displacer (BD) based MZ-interferometer \cite{Fiorentino:08}, except for the fact that the non-linear crystal (C) here is pumped continuously from both directions. Two periscopes (P1 and P2) act as half-waveplates for all three operating wavelengths. Periscope P1 works as a half-waveplate whose optic axis is at $45^{\circ}$ angle with respect to the horizontal axis; hence it rotates input horizontal polarization to vertical and vice versa. On the other side, Periscope P2 works as a half-waveplate with its optic axis at $0^{\circ}$ angle, such that it preserves the input horizontal or vertical polarization. The stability of this design comes from the two Sagnac-type loops where all the pump, signal, and idler photons are always in contact with the same optical components (except for the two BDs). So, any external drifts and fluctuations affect the pump, signal, and idler equally and get compensated by the geometry of the Sagnac itself. In fact, one may fold back the two Sagnac loops into one single loop by replacing BD2 with BD1; increasing the stability and reducing the resource cost while compromising the freedom of independent alignment or adjustment of the two BDs.

The pump beam (black line in Figure \ref{fig:Sagnac_Design}) has $+45^{\circ}$ polarization that gets divided into two parallel beams of equal intensities. The lower beam, containing horizontal polarization ($H_{p}$), is slightly delayed or advanced in time as compared to the upper beam containing vertical polarization ($V_{p}$), as the beams traverse different path-lengths and with different group velocities inside BD1 due to the birefringence property of the beam displacing material. However, if this time delay is negligibly small compared to the coherence time of the pumping laser, it only introduces a phase difference between the two pump beams which can be easily adjusted by controlling the phase of the input pump polarization. Our PPLN crystal (C) generates vertically polarized single photon pairs via type-\Romannum{0} SPDC process only for vertically polarized pump photons, $\left|V\right\rangle_{p}\rightarrow\left|V\right\rangle_{s}\left|V\right\rangle_{i}$. Hence, P1 is placed at the lower arm to convert the pump photon polarization state $\left|H\right\rangle_{p}$ to $\left|V\right\rangle_{p}$. The same periscope P1 also converts the output signal-idler pair state $\left|V\right\rangle_{s}\left|V\right\rangle_{i}$ generated from the upper pump beam into $\left|H\right\rangle_{s}\left|H\right\rangle_{i}$ state. The path of the signal (idler) photon is represented as the blue (red) dashed line. After passing P1 and P2, the upper path now has $\left|V\right\rangle_{s}\left|V\right\rangle_{i}$ two-photon state, and the lower path has $\left|H\right\rangle_{s}\left|H\right\rangle_{i}$ state, which are combined again after passing BD2. An interference filter (F) separates signal photons from the idler photons. In general, the two signal (blue) and the two idler (red) paths may not converge after BD2, as shown in Fig. \ref{fig:Sagnac_Design}. This is because the angular dispersions experience by the signal and idler photons are different from the pump photon, as the three wavelengths are significantly separated from each other. This translates to the path separation $\Delta d_{s(i)}$ and the time-delay $\Delta\tau_{s(i)}$ between the two signal (idler) wavepackets. Interestingly, one may choose the appropriate lengths and suitable birefringent materials for BD1 and BD2, such that $\Delta d_{s(i)}$ and $\Delta\tau_{s(i)}$ are negligibly small as compared to the respective coherence length and coherence time of the signal (idler) wavepacket. In that scenario, both $\left|V\right\rangle_{s}\left|V\right\rangle_{i}$ and $\left|H\right\rangle_{s}\left|H\right\rangle_{i}$ states become indistinguishable in all the degrees of freedom, leading to the entangled superposition state, $1/\sqrt{2}(\left|H\right\rangle_{s}\left|H\right\rangle_{i}+\left|V\right\rangle_{s}\left|V\right\rangle_{i})$. Similarly, we can generate the state $1/\sqrt{2}(\left|H\right\rangle_{s}\left|H\right\rangle_{i}-\left|V\right\rangle_{s}\left|V\right\rangle_{i})$ by changing the input pump polarization from $+45^{\circ}$ to $-45^{\circ}$.
\par
We use two calcite beam displacers of equal length $30~\text{mm}$ as BD1 and BD2 in our setup, entirely due to their availability at the time of the experiment. This leads to a $3.3~\text{mm}$ parallel separation of the pump beams after BD1. The time delay between the upper and lower pump beam is calculated to be $0.91~\text{ps}$, much smaller than our pump laser's coherence time ($\approx23~ps$). Hence, it can be safely assumed that this delay only introduces a phase difference between the two bidirectional pumping beams. Also, the orientation of BD2 allows partial compensation for the time delay experienced by the pump beams. Here, the pump beam traversing the extraordinary (ordinary) path in BD1 creates photon pairs that traverse the ordinary (extraordinary) path in BD2, hence, partially canceling the time delay. However, due to the significant difference between the pump and the signal, idler wavelengths, the time delay between the two signal (idler) wave packets still exists and is calculated to be $\Delta\tau_{s}=0.07~\text{ps}$ and $\Delta\tau_{i}=0.17~\text{ps}$. Interestingly, for the continuously pumped SPDC process, photon-pair generation occurs randomly in time due to the probabilistic nature of the parametric down-conversion. This means that the difference $\left|\Delta\tau_{s}-\Delta\tau_{i}\right|$ is the determining factor for qualifying temporal distinguishability rather than $\Delta\tau_{s(i)}$ themselves. From the calculation, $\left|\Delta\tau_{s}-\Delta\tau_{i}\right|=0.1~\text{ps}$, which is one order less than the coherence time of our single photons ($\approx 5~\text{ps}$), leading to a $99.8\%$ temporal overlap. Hence, the setup does not require any additional birefringent crystal for temporal wavepacket matching. The path separation values for the signal and the idler wavepackets are respectively $\Delta d_{s}=0.12~\text{mm}$ and $\Delta d_{i}=0.33~\text{mm}$. Considering the Gaussian beam diameter of $1.4~\text{mm}$ for the signal photon coupling arm and $2.1~\text{mm}$ for the idler photon coupling arm, the spatial mode overlap is calculated to be $99\%$ and $98\%$ for the signal and idler wavepackets. All the calculated values shown above are based on the theoretical descriptions shown in Ref. \cite{Lee_2021}.
\begin{figure*}
\centering    \includegraphics[width=0.8\textwidth]{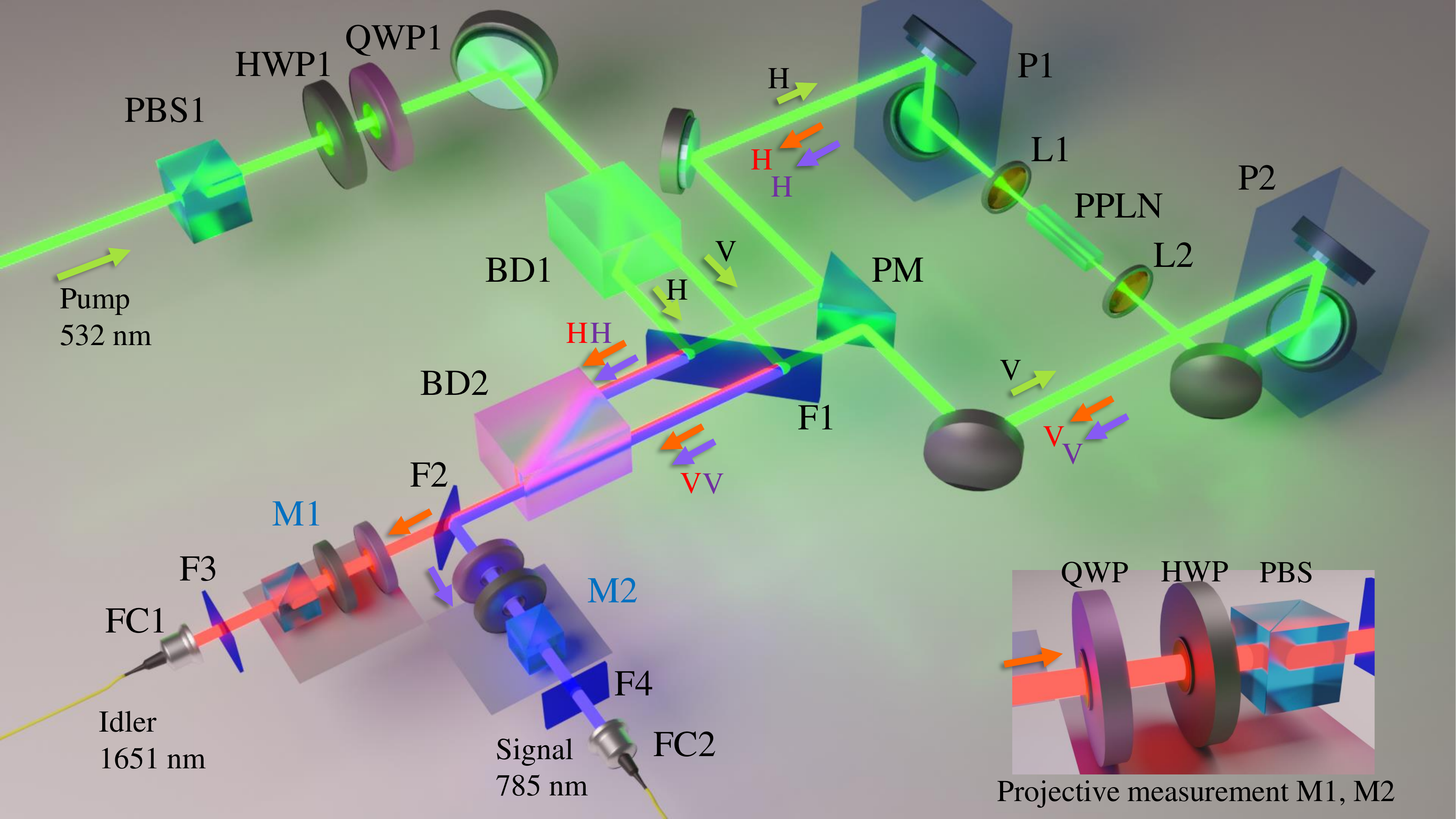}
\caption[example]
{ \label{fig:sagnac_spdc_ppln.pdf}
Schematic of the source. PBS: polarizing beamsplitter; HWP: half-waveplate; QWP: quarter-waveplate; BD1, BD2: calcite beam displacers; F1: long-pass filter (reflects $532$ nm pump beam, transmits $785$ nm signal and $1651$ nm idler photons); PM: knife-edge right-angle prism mirror; P1: periscope that rotates the input horizontal polarization to vertical and vice versa; P2: periscope that maintains the input polarization; L1, L2: focusing lens of focal length $f=50~\text{mm}$; PPLN: periodically poled lithium niobate crystal; F2: long pass dichroic mirror (transmits $1651$ nm idler photons
and reflects $785$ nm signal photons); M1, M2: setup for polarization basis measurements comprising of one QWP, one HWP, and one PBS (see insert figure); F3: long pass filter ($1500~\text{nm}$ cut-on wavelength); F4: band pass filter ($785~\text{nm}$ central wavelength, $3$ nm FWHM band); FC1: fiber coupler for coupling idler photons; FC2: fiber coupler for coupling signal photons.}
\end{figure*}

\section{Experimental setup}
In our experimental setup (Figure \ref{fig:sagnac_spdc_ppln.pdf}), a mode-locked laser (Laser Quantum opus 532) continuously pumps the non-linear crystal (PPLN) at 532 nm wavelength. The pump beam is coupled to a single-mode fiber and then collimated; to filter only the fundamental ($\text{TEM}_{0,0}$) Hermite-Gaussian spatial mode. A polarizing beamsplitter (PBS1) is placed to pass only the horizontally polarized pump beam. A half-waveplate (HWP1) and a quarter-waveplate (QWP1) convert the pump polarization into the anti-diagonally ($-45^{\circ}$) polarized beam. In principle, we can generate a variety of polarization states by rotating both HWP1 and QWP1. The general polarization state of a pump photon can be written as $\alpha \left|H\right\rangle + e^{i\phi}\beta\left|V\right\rangle$, where ideally $\alpha=\beta=1/\sqrt{2}$, and the phase value is chosen to be $\phi=180^{\circ}$. However, to compensate for the additional phase and any asymmetric loss introduced by the Sagnac interferometer (discussed in section \ref{sec:design_considerations}), we also adjust the value of $\alpha$, $\beta$, and $\phi$, as desired.
The pump beam then passes through the first calcite beam-displacer (BD1) of $30~\text{mm}$ length, where the input beam is split into two separate parallel beams based on their polarization. The vertically polarized beam maintains the original beam path, whereas the horizontally polarized beam shifts parallelly $3.3~\text{mm}$ away from the original path. Both the pump beams are reflected by a long-pass filter, F1 (Thorlabs FELH0750; reflects $532~\text{nm}$ pump beam, transmits $785~\text{nm}$ signal and $1651~\text{nm}$ idler photons), and then enter the Sagnac setup. Two pump beams first get separated by a  knife-edge right-angle prism mirror (PM) and then pass through two separate optical periscopes (P1, P2) constructed from two protected silver mirrors; before hitting the periodically-poled lithium niobate (PPLN) crystal. Here, one periscope (P1) rotates the polarization of the input pump beam by $90^{\circ}$ (horizontal polarization becomes vertical polarization and vice versa), and the other one (P2) does not change the polarization but is used only for changing the height of the beam such that both pump beams have equal heights when they hit the PPLN crystal.
\par
The PPLN crystal is quasi-phase-matched to produce single photon pairs via the type-\Romannum{0} SPDC process. Our crystal has a dimension of $(0.5~\text{mm}\times 0.5~\text{mm}\times 20~\text{mm})$ and a poling period of $7.71~\mu \text{m}$. To achieve optimal quasi-phase matching, the crystal is put inside an oven whose temperature is stabilized at $62^{\circ} C$ using a PID controller. In this condition, a vertically polarized pump photon (532 nm wavelength) can generate both vertically-polarized signal (785 nm) and idler (1651 nm) photons in a collinear configuration. To optimize the photon pair generation rate, both the pump beams are equally focused onto the center of the PPLN crystal and then collimated again using two identical lenses (L1, L2). 
\par
The generated photon pairs from the horizontally (vertically) polarized pump beam follow the same path as the vertically (horizontally) polarized input pump beam inside the Sagnac setup. When they reach the long-pass filter (F1), the pump beams get reflected again, but the photon pairs pass through it, and both the photon paths merge into a single path after the second beam-displacer (BD2) of $30~\text{mm}$ length. A long pass dichroic mirror, F2 (Thorlabs DLMP1180; transmits $1651~\text{nm}$ idler photons and reflects $785~\text{nm}$ signal photons) is used to separate the signal photon from the idler photon and send them to two separate fiber couplers (FC1, FC2) connected to single-photon detectors. For the signal photons, we use a silicon avalanche detector (Perkin Elmer SPCM-AQRH-14-FC) with a detection efficiency of $50\%$, the dark count rate of $100~\text{cps}$ and $32~\text{ns}$ deadtime. For the idler photon, a superconducting nanowire detector (Scontel SNSPD) is used, whose efficiency is around $45\%$ for the $1651~\text{nm}$ wavelength. Both detectors are connected to a time-tagging module (quTAG) which performs the time-correlated single photon counting (TCSPC) and coincidence measurement for a given coincidence window of $3~\text{ns}$.

To understand how the entangled two-photon state, $\left|\Psi^{-}\right\rangle=(\left|HH\right\rangle - \left|VV\right\rangle)/\sqrt{2}$ is generated inside the Sagnac setup; we have to focus on the individual pump beams separately; first, the horizontally polarized pump beam (H beam) and then the vertically polarized beam (V beam). The H beam passes through the periscope P1 and converts to the vertically polarized beam, which upon interacting with the PPLN crystal, creates both vertically polarized signal and idler photons ($\left|V\right\rangle_{pump}\rightarrow \left|V\right\rangle_{signal}\left|V\right\rangle_{idler}$). These photon pairs pass through periscope P2, which maintains their polarization and only changes the path's height. These photon pairs maintain their original path while passing through the beam-displacer BD2. On the other hand, the V beam passes through the periscope P2, which maintains the polarization but lowers the beam height. The V beam interacts with the PPLN crystal and produces pair of single photons with vertical polarization ($\left|V\right\rangle_{pump}\rightarrow \left|V\right\rangle_{signal}\left|V\right\rangle_{idler}$). Both vertically polarized signal and idler photons are then converted to horizontally polarized photons ($\left|V\right\rangle_{signal}\left|V\right\rangle_{idler}\rightarrow \left|H\right\rangle_{signal}\left|H\right\rangle_{idler}$) after passing the periscope P1. These photons follow the exact path of the H pump beam in a reverse direction. After passing BD2, the path of the photon pairs merged with the path of the photon pairs from the H pump beam. As the SPDC process is a probabilistic process with a small probability of photon-pair generation ($\approx 10^{-10}$ ), it has a negligibly small probability for both pump beams to create photon pairs simultaneously. Hence, at any time instance, there is either an $\left|H\right\rangle_{signal}\left|H\right\rangle_{idler}$ state or $\left|V\right\rangle_{signal}\left|V\right\rangle_{idler}$ state after BD2. If the two states are entirely indistinguishable in all other degrees of freedom (spatial, temporal, spectral, etc.), according to the quantum mechanical superposition principle, we can write the state as $(\left|H\right\rangle\left|H\right\rangle \pm\left|V\right\rangle\left|V\right\rangle)/\sqrt{2}$. The phase of the input polarization state of the pump beam determines the plus or minus sign.

\section{Results}
\begin{figure*}
\centering    
\includegraphics[width=0.9\textwidth]{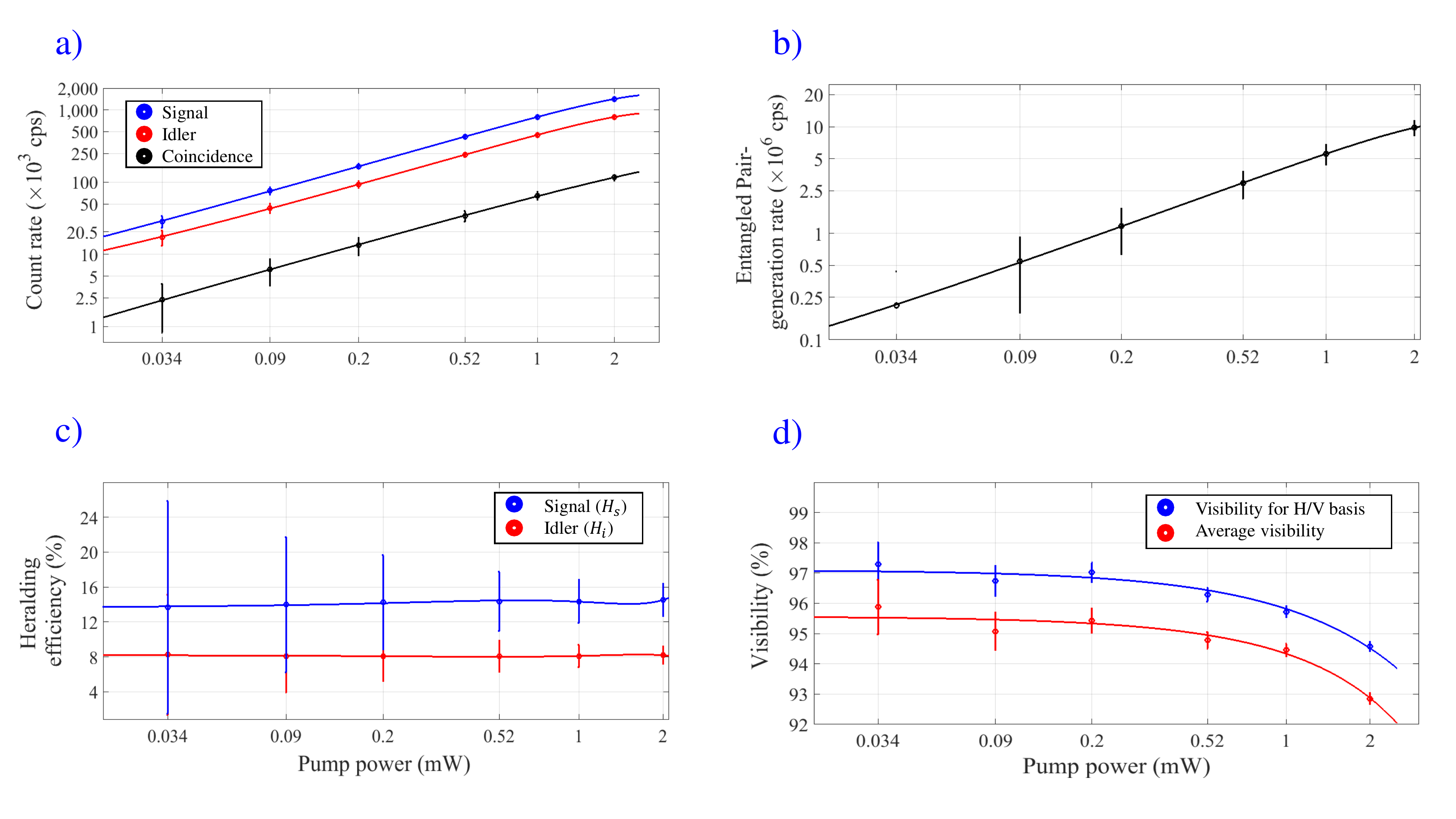}
\caption[example]
{ \label{fig:Sagnac_all4.pdf}
Performance of the source. \textbf{a)} plot of entangled photon-pair coincidence rate (black circle) and signal (blue circle), idler (red circle) count rates for different values of pump power. Each data point is the average value over 10 seconds of collection time and over 3 polarization-basis measurement settings; rectilinear ($H/V$), diagonal ($D/A$), and circular ($R/L$). The error bar represents the standard deviation of each data point with the assumption of Poissonian statistics. Plot of \textbf{b)} entangled pair-generation rate, \textbf{c)} heralding efficiencies for signal and idler arms, and \textbf{d)} visibility of the polarization-entangled photon pairs against the pump power value. The blue circle in \textbf{d} represents the measured visibility in the rectilinear ($H/V$) basis, and the red circle represents the average visibility measured over 3 separate polarization bases ($H/V$, $D/A$, and $R/L$). We show second-order polynomial fitting for every dataset in Figures \textbf{a}, \textbf{b}, and \textbf{c}, whereas exponential fitting is applied for \textbf{d}. To cover the broad range of the pump power values, logarithmic visualization is applied for both the x and y-axis of the plot.}
\end{figure*}

We measure the signal and the idler count rate along with the raw coincidence rate for different pump power values in our Sagnac setup. For this purpose, six different pump power is chosen; $34~\mu \text{W}$, $90~\mu \text{W}$, $200~\mu \text{W}$, $520~\mu \text{W}$, $1~\text{mW}$, and $2~\text{mW}$. The quoted power is the total power sent to the PPLN crystal, so the pump power on each side of the crystal (H beam or V beam in Figure \ref{fig:sagnac_spdc_ppln.pdf}) is roughly half of the quoted power. All the measurements are performed for the optimal collection efficiency of the fiber couplers.
 Figure \ref {fig:Sagnac_all4.pdf} shows all the experimental data points along with the numerical fitting to the data.


\par
Our Sagnac setup is creating the entangled state $(\left|H\right\rangle\left|H\right\rangle - \left|V\right\rangle\left|V\right\rangle)/\sqrt{2}$, which also can be represented as $(\left|D\right\rangle\left|A\right\rangle + \left|A\right\rangle\left|D\right\rangle)/\sqrt{2}$ and $(\left|R\right\rangle\left|R\right\rangle + \left|L\right\rangle\left|L\right\rangle)/\sqrt{2}$ state. Here, $\left|D(A)\right\rangle=\left|H\right\rangle\pm\left|V\right\rangle$, and $\left|R(L)\right\rangle=\left|H\right\rangle\pm i\left|V\right\rangle$. If one considers only one output arm, say the signal arm (or idler arm), the state can be written as a maximally mixed state $(\left|H\right\rangle\left\langle H\right|+\left|V\right\rangle\left\langle V\right|)/2=(\left|D\right\rangle\left\langle D\right|+\left|A\right\rangle\left\langle A\right|)/2=(\left|R\right\rangle\left\langle R\right|+\left|L\right\rangle\left\langle L\right|)/2$. Hence, we measure the sum of the count rate for both projections $\left|H\right\rangle\left\langle H\right|$ and $\left|V\right\rangle\left\langle V\right|$ in the signal (idler) arm. Similarly, we also measure the sum of the count rate for both $\left|D\right\rangle\left\langle D\right|$ and $\left|A\right\rangle\left\langle A\right|$ projection, and again for $\left|R\right\rangle\left\langle R\right|$ and $\left|L\right\rangle\left\langle L\right|$ projections. The quoted count rate value in Figure \ref {fig:Sagnac_all4.pdf}-a for both signal and idler is the average value of these three count rates. Similarly, for the coincidence count rate shown in Figure \ref {fig:Sagnac_all4.pdf}-a, we take an average of the three coincidence rates; the sum of coincidence rates for both $\left|H\right\rangle\left|H\right\rangle\left\langle H\right|\left\langle H\right|$ and $\left|V\right\rangle\left|V\right\rangle\left\langle V\right|\left\langle V\right|$ projections, the sum of coincidence rates for both $\left|D\right\rangle\left|A\right\rangle\left\langle D\right|\left\langle A\right|$ and $\left|A\right\rangle\left|D\right\rangle\left\langle A\right|\left\langle D\right|$ projections, and the sum of coincidence rates for both $\left|R\right\rangle\left|R\right\rangle\left\langle R\right|\left\langle R\right|$ and $\left|L\right\rangle\left|L\right\rangle\left\langle L\right|\left\langle L\right|$ projections. Also, the quoted coincidence rate in Figure \ref {fig:Sagnac_all4.pdf}-a is corrected for the accidental coincidence rate, which is calculated as the (signal count rate $\times$ idler count rate $\times$ coincidence window). This accidental coincidence rate is subtracted from the raw coincidence rate for each data point. Here, we consider a coincidence window of $3~\text{ns}$ for every measurement.
\par
We also calculate the heralding efficiencies (Figure \ref{fig:Sagnac_all4.pdf}-c) and the entangled pair-generation rate (Figure \ref{fig:Sagnac_all4.pdf}-b) from the experimental data. Heralding efficiency for signal (idler) is calculated using the formula $H_{s(i)}=C/S_{s(i)}$, where $C$ is the coincidence rate and $S_{s(i)}$ is the signal (idler) count rate. Entangled pair-generation rate ($PGR$) is calculated as $PGR=(S_{s}\times S_{i})/C$. The error bar for each data point shown in Figure \ref{fig:Sagnac_all4.pdf} is calculated as $\pm\sigma$, where $\sigma=\sqrt{N}$ considering the Poissonian distribution for the photon source. We employ a second-order polynomial fitting for all the datasets in Figure \ref{fig:Sagnac_all4.pdf}-a, \ref{fig:Sagnac_all4.pdf}-b, \ref{fig:Sagnac_all4.pdf}-c, which closely matches our experimental data. We observe a slight saturation effect in the single count rates and the coincidence rate for high pump power ($\approx 2 mW$), which is due to the multi-photon pair generation as well as the saturation effect of the single-photon detectors \cite{steinlechner2012high}; according to our speculation.


We measure the polarization-visibility of our entangled photon pairs for different pump power values. For the measurement of the visibility in ($H/V$) basis, we measure all four projection values; $\left|H\right\rangle\left|H\right\rangle\left\langle H\right|\left\langle H\right|$, $\left|H\right\rangle\left|V\right\rangle\left\langle H\right|\left\langle V\right|$, $\left|V\right\rangle\left|H\right\rangle\left\langle V\right|\left\langle H\right|$, and $\left|V\right\rangle\left|V\right\rangle\left\langle V\right|\left\langle V\right|$. Polarization-visibility ($PV$) in the ($H/V$) basis is then calculated as,
\begin{equation*}
\resizebox{0.45\textwidth}{!}{$PV_{(H/V)}$=$\frac{C_{\left|H\right\rangle\left|H\right\rangle\left\langle H\right|\left\langle H\right|}-C_{\left|H\right\rangle\left|V\right\rangle\left\langle H\right|\left\langle V\right|}-C_{\left|V\right\rangle\left|H\right\rangle\left\langle V\right|\left\langle H\right|}+C_{\left|V\right\rangle\left|V\right\rangle\left\langle V\right|\left\langle V\right|}}{C_{\left|H\right\rangle\left|H\right\rangle\left\langle H\right|\left\langle H\right|}+C_{\left|H\right\rangle\left|V\right\rangle\left\langle H\right|\left\langle V\right|}+C_{\left|V\right\rangle\left|H\right\rangle\left\langle V\right|\left\langle H\right|}+C_{\left|V\right\rangle\left|V\right\rangle\left\langle V\right|\left\langle V\right|}}$}
\end{equation*}
Here, $C_{\left|i\right\rangle\left|j\right\rangle\left\langle i\right|\left\langle j\right|}$ is the coincidence count rate for the projection measurement $\left|i\right\rangle\left|j\right\rangle\left\langle i\right|\left\langle j\right|$. We plot the behavior of $PV_{(H/V)}$ with increasing pump power in Figure \ref{fig:Sagnac_all4.pdf}-d (blue-colored data points). As one can observe, the visibility value slowly drops down with higher pump power. We correlate this behavior to the higher multi-photon generation probability at high pumping power, also observed in \cite{schneeloch2019introduction,steinlechner2012high}. In addition to the pump power, the maximum achievable visibility is also limited by the size of the coincidence window and other experimental imperfections, such as the polarization extinction ratios of the polarization analyzers, the alignment of the Sagnac setup and its phase stability, coupling efficiency of the fiber couplers, etc. A lower value of the coincidence window enhances the visibility value. However, the selection of a suitable coincidence window size is also limited by the timing jitter of the single-photon detectors.

We also plot the average visibility value in Figure \ref{fig:Sagnac_all4.pdf}-d (red colored data points), which is the average of the three visibility values measured in three conjugate bases; $PV_{H/V}$, $PV_{D/A}$, and $PV_{R/L}$. We observe a slight lowering of the average visibility as compared to $PV_{H/V}$. For example, at $0.034~\mu \text{W}$ pump power, experimentally measured $PV_{H/V}$ value is $97.29\%$ while the average visibility is $95.88\%$. This lowering in the average visibility also qualifies the degree of entanglement between the photon pairs. For an ideal experimental condition, maximal entanglement should produce equal visibility on any orthogonal polarization basis. However, the high value of the average visibility from our source signifies a good quality of entanglement. 
\par
As one may observe, the brightness of our non-degenerate entangled photon source is very high. This is partly due to the usage of PPLN crystal as the non-linear medium, which has a high pair-generation rate due to its large non-linear ($\chi^2$) coefficient. We detect entangled photon pairs of high brightness even at a very low pump power of $34 ~\mu \text{W}$. At this pump power value, we measure the average coincidence rate of $(2.37\pm 0.05)\times 10^{3}$ cps. Hence, we report our average detected coincidence rate as $(6.96\pm 0.03)\times 10^{4}$ cps/mW. The entangled pair generation rate (PGR) is calculated to be $(6.17\pm 0.04)\times 10^{6}$ cps/mW, using the formula, $PGR=(S_{s}\times S_{i})/C$, as mentioned before. We measure the polarization correlation between the signal and idler photons at this low pump power value. For this purpose, we place a polarization basis measurement setup consisting of a half-waveplate and a polarizing beamsplitter (PBS), in the two output arms of the source (see Figure \ref{fig:sagnac_spdc_ppln.pdf}). By rotating the half-waveplate angles from $0^{\circ}$ to $180^{\circ}$, we can project the whole $360^{\circ}$ polarization direction in the transmitting arm of the PBS. In our polarization correlation measurement, we fix the projection on the signal photons to four different directions; horizontal (H, $0^{\circ}$), vertical (V, $90^{\circ}$), diagonal (D, $45^{\circ}$), and anti-diagonal (A, $135^{\circ}$). For each fixed projection measurement on the signal photons, the projection angle on the idler photons is continuously rotated from $0^{\circ}$ to $360^{\circ}$ with the interval of $12^{\circ}$, and the average raw coincidence count rate is measured for 10 seconds of data collection. We observe a sinusoidal pattern in the raw coincidence count rate for all four fixed projection angles on the signal photons, with high visibility, $V=(\text{Max coincidence}-\text{Min coincidence})/(\text{Max coincidence}+\text{Min coincidence})$. We fit the experimental data with the sinusoidal curve. From the fitted curves, we calculate the visibility as $96.49\pm 00.76\%$ (H), $95.26\pm 00.89\%$ (V), $96.01\pm 00.81\%$ (D), and $96.27\pm 00.79\%$ (A). We also measure the Bell-CHSH inequality value, $S= \left\langle AB^{\phantom{'}} \right\rangle - \left\langle AB^{'}\right\rangle - \left\langle A^{'}B\right\rangle - \left\langle A^{'}B^{'}\right\rangle$, where $A$ and $A^{'}$ are the two projection basis measurement on the signal photons with polarization angle $0^{\circ}$ and $45^{\circ}$ respectively, and $B$, $B^{'}$ are the two projection basis measurement on the idler photons with polarization angle $22.5^{\circ}$ and $67.5^{\circ}$ respectively \cite{clauser1969proposed}. We observe a Bell-CHSH value of $S=2.71\pm 0.06$, which violates the local-real bound of $S\leq 2$, by $11.8$ standard deviation.

\begin{figure}
\centering
\includegraphics[height=5cm]{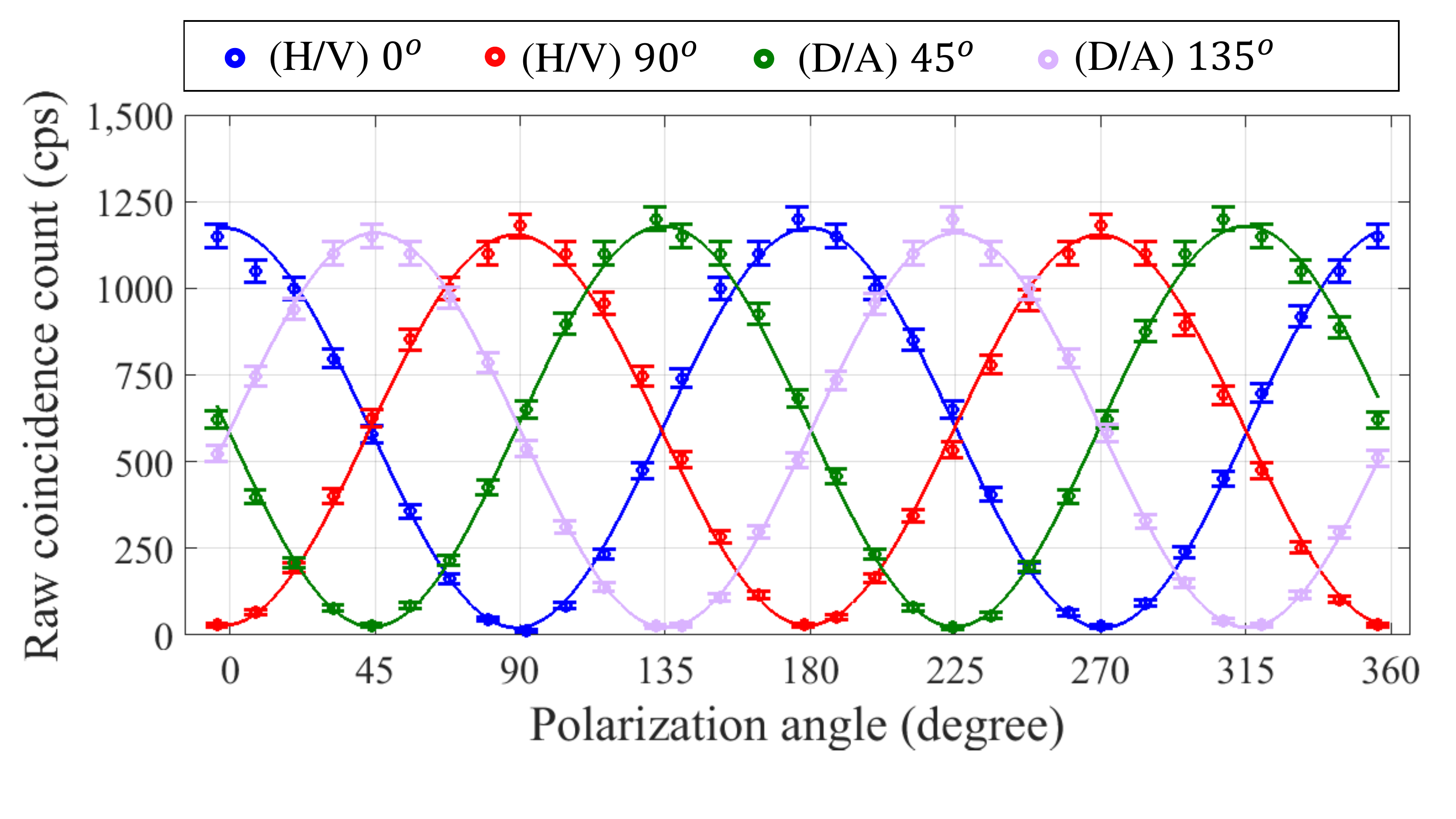}
\caption{The degree of polarization-entanglement is shown by plotting the raw coincidence counts while the signal polarization basis is fixed and the idler polarization analyzer is rotated by a full $2\pi$ rotation. The polarization measurement basis of the signal photon is fixed for 4 directions; horizontal (H, blue circle), vertical (V, red circle), diagonal (D, green circle), and anti-diagonal (A, pink circle). The colored circles are the experimental raw coincidence count rates (without any background correction) averaged over 10 seconds of data collection. Error bars are the standard deviations calculated with the assumption of Poissonian statistics. Sinusoidal fitting is applied on the 4 datasets; for the lowest chi-squared value.}
\label{fig:Sagnac_4polarization.pdf}
\end{figure}
\par
\begin{figure}
\begin{center}
\includegraphics[height=5cm]{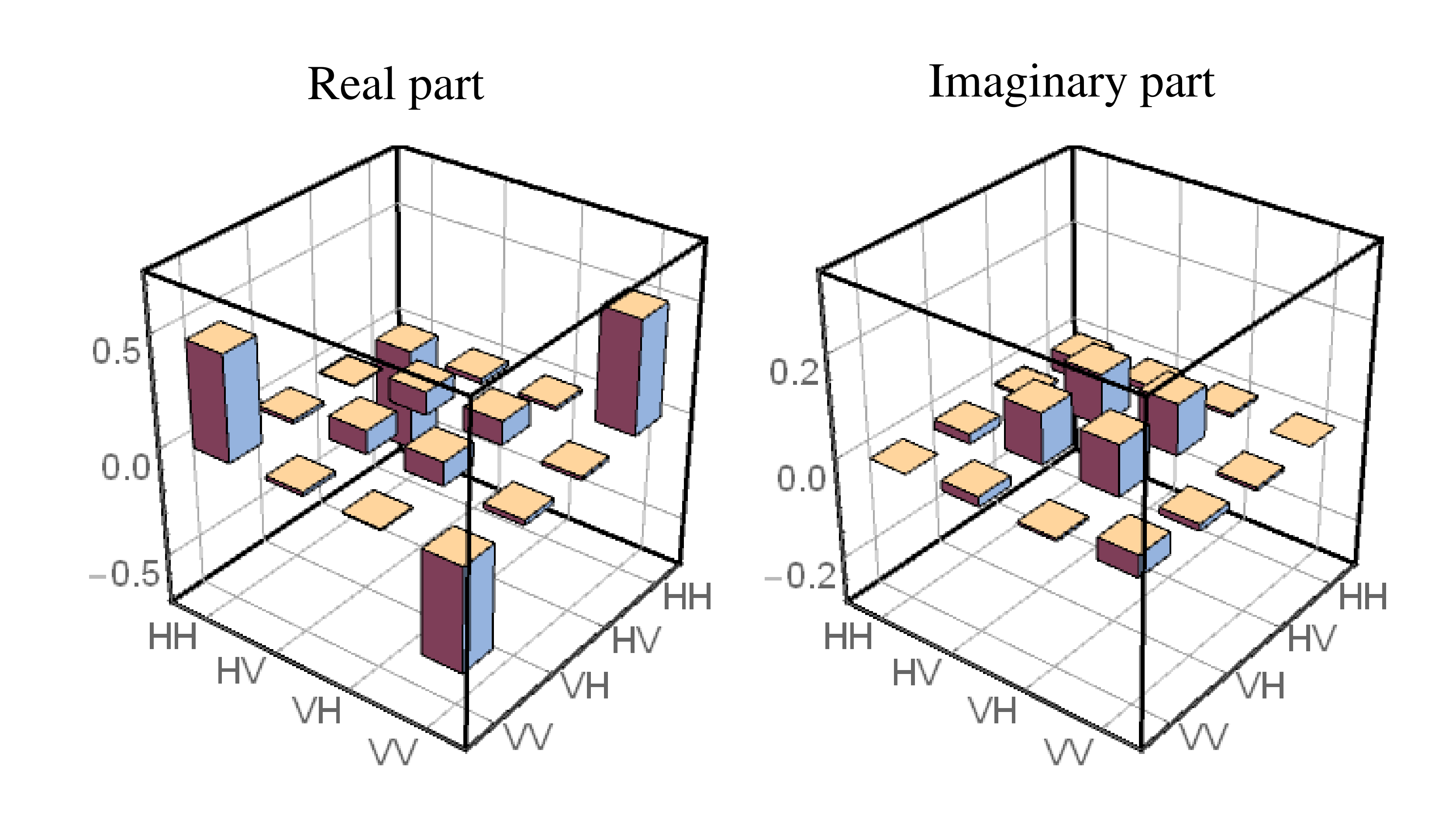}
\end{center}
\caption[example]
{ \label{fig:density matrix}
The real part (left) and the imaginary part (right) of the reconstructed density matrix, as obtained from the maximum likelihood estimation (MLE) process.}
\end{figure}

We measure the tomography of the generated photon-pair state; to quantify the degree of entanglement as well as the closeness (fidelity) of the output state to the desired entangled state, $\left|\Phi^{-}\right\rangle = (\left|H\right\rangle\left|H\right\rangle - \left|V\right\rangle\left|V\right\rangle)/\sqrt{2} $. For this purpose, we implement the tomography setup in the two output arms, shown as projective measurements M1 and M2 in Figure \ref{fig:sagnac_spdc_ppln.pdf}. We measure the coincidence rate for 16 projective measurements; $H_{s}H_{i}$, $H_{s}V_{i}$, $V_{s}V_{i}$, $V_{s}H_{i}$, $L_{s}H_{i}$, $L_{s}V_{i}$, $D_{s}V_{i}$, $D_{s}H_{i}$, $D_{s}L_{i}$, $D_{s}D_{i}$, $L_{s}D_{i}$, $H_{s}D_{i}$, $V_{s}D_{i}$, $V_{s}R_{i}$, $L_{s}D_{i}$, $H_{s}R_{i}$ projections. Here, $X_{s}Y_{i}$ represents the projective operation $\left|X\right\rangle_{s}\left|Y\right\rangle_{i}\left\langle X\right|_{s}\left\langle Y\right|_{i}$. For each projective measurement, the coincidence rate is measured as an average of data collected over 10 seconds, while the pump power is only $34~\mu W$. From these datasets, the density matrix of the photon-pair state is reconstructed using the maximum likelihood estimation (MLE) process \cite{james2001measurement}. We find that for our reconstructed density matrix, the concurrence is $94.68\pm 0.20$, purity of the state is $94.40\pm 0.25$, and the fidelity is $96.72\pm 0.01$ concerning the $\left|\Phi^{-}\right\rangle$ state. Fig. \ref{fig:density matrix} shows the real and the imaginary part of the reconstructed two-photon density matrix.

\section{Conclusion}
We developed a Sagnac-type source for non-degenerate polarization-entangled photon pairs. The signal photon wavelength ($785~\text{nm}$) is vastly different from the idler photon wavelength ($1651~\text{nm}$). Our source satisfies all the required conditions of high performance, low resource requirement, stability, and adaptability. We achieved a significantly high brightness of $(6.17\pm0.04)\times10^{6}$ entangled pairs/sec/mW, whereas the fidelity is $(96.72\pm 0.01)\%$ with respect to the desired two-photon entangled state, $\left|\Psi^{-}\right\rangle=(\left|HH\right\rangle - \left|VV\right\rangle)/\sqrt{2}$. We used only off-the-shelf optical components, and the source design is highly compact, which reduces the total cost and weight. The Sagnac-type design helps attain long-term phase stability without using any active phase stabilization method, further improving the resource requirements. Although we demonstrated the performance of our source with continuous wave (CW) pumping, the source can perform equally with pulsed pumping conditions. In that scenario, one has to include two compensating birefringent crystals, one before and one after the Sagnac loop, to compensate for temporal mismatch. Our source design can be adapted for type-\Romannum{1} and type-\Romannum{2} phase-matching conditions with minimal modification (only one additional half-waveplate or another beam displacer is required \cite{Horn:19}). One may also miniaturize the whole setup and fit it inside a smaller dimension, further increasing the stability. For this purpose, one may remove the two focusing lenses (L1, L2) and place them outside the Sagnac loop. 
\par
Our non-degenerate polarization-entangled photon-pair source with high brightness and fidelity is ideal for application in satellite-to-ground-based QKD links. Also, using PPLN crystal for the non-degenerate SPDC process reduces the line width of the generated photon pairs, as compared to the conventional BBO crystal. With additional cavity-based spectral filtering, the line width of the signal photons can be reduced even further to match the atomic transition line width. In that scenario, our source can be used for a quantum-memory-based communication channel. Ultrabright photon pair sources such as the one described here can perform even with tight spectral filtering.
The performance of the source can be further modified with some additional changes. One may use $\alpha$-BBO beam displacing crystals instead of calcite to reduce further the temporal and spatial distinguishability between the two wavepackets coming from the bidirectional pumping. Also, different lengths of beam displacing crystals (BD1, BD2) can further reduce $\Delta d_{s(i)}$ and $\Delta\tau_{s(i)}$ values, resulting in higher polarization visibility.

\medskip

\noindent\textbf{Funding Information.} The authors acknowledge financial support by the Foundation for Polish Science (FNP) (project First Team co-financed by the European Union under the European Regional Development Fund, grant no. First Team/2017-3/20).

\medskip

\noindent\textbf{Disclosures.} The authors declare no conflicts of interest.

\bibliography{report} 

\begin{thebibliography}{27}%
\makeatletter
\providecommand \@ifxundefined [1]{%
 \@ifx{#1\undefined}
}%
\providecommand \@ifnum [1]{%
 \ifnum #1\expandafter \@firstoftwo
 \else \expandafter \@secondoftwo
 \fi
}%
\providecommand \@ifx [1]{%
 \ifx #1\expandafter \@firstoftwo
 \else \expandafter \@secondoftwo
 \fi
}%
\providecommand \natexlab [1]{#1}%
\providecommand \enquote  [1]{``#1''}%
\providecommand \bibnamefont  [1]{#1}%
\providecommand \bibfnamefont [1]{#1}%
\providecommand \citenamefont [1]{#1}%
\providecommand \href@noop [0]{\@secondoftwo}%
\providecommand \href [0]{\begingroup \@sanitize@url \@href}%
\providecommand \@href[1]{\@@startlink{#1}\@@href}%
\providecommand \@@href[1]{\endgroup#1\@@endlink}%
\providecommand \@sanitize@url [0]{\catcode `\\12\catcode `\$12\catcode
  `\&12\catcode `\#12\catcode `\^12\catcode `\_12\catcode `\%12\relax}%
\providecommand \@@startlink[1]{}%
\providecommand \@@endlink[0]{}%
\providecommand \url  [0]{\begingroup\@sanitize@url \@url }%
\providecommand \@url [1]{\endgroup\@href {#1}{\urlprefix }}%
\providecommand \urlprefix  [0]{URL }%
\providecommand \Eprint [0]{\href }%
\providecommand \doibase [0]{http://dx.doi.org/}%
\providecommand \selectlanguage [0]{\@gobble}%
\providecommand \bibinfo  [0]{\@secondoftwo}%
\providecommand \bibfield  [0]{\@secondoftwo}%
\providecommand \translation [1]{[#1]}%
\providecommand \BibitemOpen [0]{}%
\providecommand \bibitemStop [0]{}%
\providecommand \bibitemNoStop [0]{.\EOS\space}%
\providecommand \EOS [0]{\spacefactor3000\relax}%
\providecommand \BibitemShut  [1]{\csname bibitem#1\endcsname}%
\let\auto@bib@innerbib\@empty
\bibitem [{\citenamefont {Horodecki}\ \emph {et~al.}(2009)\citenamefont
  {Horodecki}, \citenamefont {Horodecki}, \citenamefont {Horodecki},\ and\
  \citenamefont {Horodecki}}]{horodecki2009quantum}%
  \BibitemOpen
  \bibfield  {author} {\bibinfo {author} {\bibfnamefont {R.}~\bibnamefont
  {Horodecki}}, \bibinfo {author} {\bibfnamefont {P.}~\bibnamefont
  {Horodecki}}, \bibinfo {author} {\bibfnamefont {M.}~\bibnamefont
  {Horodecki}}, \ and\ \bibinfo {author} {\bibfnamefont {K.}~\bibnamefont
  {Horodecki}},\ }\href@noop {} {\bibfield  {journal} {\bibinfo  {journal}
  {Reviews of modern physics}\ }\textbf {\bibinfo {volume} {81}},\ \bibinfo
  {pages} {865} (\bibinfo {year} {2009})}\BibitemShut {NoStop}%
\bibitem [{\citenamefont {Bennett}\ \emph {et~al.}(1993)\citenamefont
  {Bennett}, \citenamefont {Brassard}, \citenamefont {Cr\'epeau}, \citenamefont
  {Jozsa}, \citenamefont {Peres},\ and\ \citenamefont
  {Wootters}}]{PhysRevLett.70.1895}%
  \BibitemOpen
  \bibfield  {author} {\bibinfo {author} {\bibfnamefont {C.~H.}\ \bibnamefont
  {Bennett}}, \bibinfo {author} {\bibfnamefont {G.}~\bibnamefont {Brassard}},
  \bibinfo {author} {\bibfnamefont {C.}~\bibnamefont {Cr\'epeau}}, \bibinfo
  {author} {\bibfnamefont {R.}~\bibnamefont {Jozsa}}, \bibinfo {author}
  {\bibfnamefont {A.}~\bibnamefont {Peres}}, \ and\ \bibinfo {author}
  {\bibfnamefont {W.~K.}\ \bibnamefont {Wootters}},\ }\href {\doibase
  10.1103/PhysRevLett.70.1895} {\bibfield  {journal} {\bibinfo  {journal}
  {Phys. Rev. Lett.}\ }\textbf {\bibinfo {volume} {70}},\ \bibinfo {pages}
  {1895} (\bibinfo {year} {1993})}\BibitemShut {NoStop}%
\bibitem [{\citenamefont {Bennett}\ and\ \citenamefont
  {Wiesner}(1992)}]{PhysRevLett.69.2881}%
  \BibitemOpen
  \bibfield  {author} {\bibinfo {author} {\bibfnamefont {C.~H.}\ \bibnamefont
  {Bennett}}\ and\ \bibinfo {author} {\bibfnamefont {S.~J.}\ \bibnamefont
  {Wiesner}},\ }\href {\doibase 10.1103/PhysRevLett.69.2881} {\bibfield
  {journal} {\bibinfo  {journal} {Phys. Rev. Lett.}\ }\textbf {\bibinfo
  {volume} {69}},\ \bibinfo {pages} {2881} (\bibinfo {year}
  {1992})}\BibitemShut {NoStop}%
\bibitem [{\citenamefont {Ekert}(1991)}]{PhysRevLett.67.661}%
  \BibitemOpen
  \bibfield  {author} {\bibinfo {author} {\bibfnamefont {A.~K.}\ \bibnamefont
  {Ekert}},\ }\href {\doibase 10.1103/PhysRevLett.67.661} {\bibfield  {journal}
  {\bibinfo  {journal} {Phys. Rev. Lett.}\ }\textbf {\bibinfo {volume} {67}},\
  \bibinfo {pages} {661} (\bibinfo {year} {1991})}\BibitemShut {NoStop}%
\bibitem [{\citenamefont {Bennett}\ \emph {et~al.}(1992)\citenamefont
  {Bennett}, \citenamefont {Brassard},\ and\ \citenamefont
  {Mermin}}]{PhysRevLett.68.557}%
  \BibitemOpen
  \bibfield  {author} {\bibinfo {author} {\bibfnamefont {C.~H.}\ \bibnamefont
  {Bennett}}, \bibinfo {author} {\bibfnamefont {G.}~\bibnamefont {Brassard}}, \
  and\ \bibinfo {author} {\bibfnamefont {N.~D.}\ \bibnamefont {Mermin}},\
  }\href {\doibase 10.1103/PhysRevLett.68.557} {\bibfield  {journal} {\bibinfo
  {journal} {Phys. Rev. Lett.}\ }\textbf {\bibinfo {volume} {68}},\ \bibinfo
  {pages} {557} (\bibinfo {year} {1992})}\BibitemShut {NoStop}%
\bibitem [{\citenamefont {Steinlechner}\ \emph {et~al.}(2013)\citenamefont
  {Steinlechner}, \citenamefont {Ramelow}, \citenamefont {Jofre}, \citenamefont
  {Gilaberte}, \citenamefont {Jennewein}, \citenamefont {Torres}, \citenamefont
  {Mitchell},\ and\ \citenamefont {Pruneri}}]{steinlechner2013phase}%
  \BibitemOpen
  \bibfield  {author} {\bibinfo {author} {\bibfnamefont {F.}~\bibnamefont
  {Steinlechner}}, \bibinfo {author} {\bibfnamefont {S.}~\bibnamefont
  {Ramelow}}, \bibinfo {author} {\bibfnamefont {M.}~\bibnamefont {Jofre}},
  \bibinfo {author} {\bibfnamefont {M.}~\bibnamefont {Gilaberte}}, \bibinfo
  {author} {\bibfnamefont {T.}~\bibnamefont {Jennewein}}, \bibinfo {author}
  {\bibfnamefont {J.~P.}\ \bibnamefont {Torres}}, \bibinfo {author}
  {\bibfnamefont {M.~W.}\ \bibnamefont {Mitchell}}, \ and\ \bibinfo {author}
  {\bibfnamefont {V.}~\bibnamefont {Pruneri}},\ }\href@noop {} {\bibfield
  {journal} {\bibinfo  {journal} {Optics express}\ }\textbf {\bibinfo {volume}
  {21}},\ \bibinfo {pages} {11943} (\bibinfo {year} {2013})}\BibitemShut
  {NoStop}%
\bibitem [{\citenamefont {Marcikic}\ \emph {et~al.}(2004)\citenamefont
  {Marcikic}, \citenamefont {De~Riedmatten}, \citenamefont {Tittel},
  \citenamefont {Zbinden}, \citenamefont {Legr{\'e}},\ and\ \citenamefont
  {Gisin}}]{marcikic2004distribution}%
  \BibitemOpen
  \bibfield  {author} {\bibinfo {author} {\bibfnamefont {I.}~\bibnamefont
  {Marcikic}}, \bibinfo {author} {\bibfnamefont {H.}~\bibnamefont
  {De~Riedmatten}}, \bibinfo {author} {\bibfnamefont {W.}~\bibnamefont
  {Tittel}}, \bibinfo {author} {\bibfnamefont {H.}~\bibnamefont {Zbinden}},
  \bibinfo {author} {\bibfnamefont {M.}~\bibnamefont {Legr{\'e}}}, \ and\
  \bibinfo {author} {\bibfnamefont {N.}~\bibnamefont {Gisin}},\ }\href@noop {}
  {\bibfield  {journal} {\bibinfo  {journal} {Physical Review Letters}\
  }\textbf {\bibinfo {volume} {93}},\ \bibinfo {pages} {180502} (\bibinfo
  {year} {2004})}\BibitemShut {NoStop}%
\bibitem [{\citenamefont {Walborn}\ \emph {et~al.}(2010)\citenamefont
  {Walborn}, \citenamefont {Monken}, \citenamefont {P{\'a}dua},\ and\
  \citenamefont {Ribeiro}}]{walborn2010spatial}%
  \BibitemOpen
  \bibfield  {author} {\bibinfo {author} {\bibfnamefont {S.~P.}\ \bibnamefont
  {Walborn}}, \bibinfo {author} {\bibfnamefont {C.}~\bibnamefont {Monken}},
  \bibinfo {author} {\bibfnamefont {S.}~\bibnamefont {P{\'a}dua}}, \ and\
  \bibinfo {author} {\bibfnamefont {P.~S.}\ \bibnamefont {Ribeiro}},\
  }\href@noop {} {\bibfield  {journal} {\bibinfo  {journal} {Physics Reports}\
  }\textbf {\bibinfo {volume} {495}},\ \bibinfo {pages} {87} (\bibinfo {year}
  {2010})}\BibitemShut {NoStop}%
\bibitem [{\citenamefont {Anwar}\ \emph {et~al.}(2021)\citenamefont {Anwar},
  \citenamefont {Perumangatt}, \citenamefont {Steinlechner}, \citenamefont
  {Jennewein},\ and\ \citenamefont {Ling}}]{doi:10.1063/5.0023103}%
  \BibitemOpen
  \bibfield  {author} {\bibinfo {author} {\bibfnamefont {A.}~\bibnamefont
  {Anwar}}, \bibinfo {author} {\bibfnamefont {C.}~\bibnamefont {Perumangatt}},
  \bibinfo {author} {\bibfnamefont {F.}~\bibnamefont {Steinlechner}}, \bibinfo
  {author} {\bibfnamefont {T.}~\bibnamefont {Jennewein}}, \ and\ \bibinfo
  {author} {\bibfnamefont {A.}~\bibnamefont {Ling}},\ }\href {\doibase
  10.1063/5.0023103} {\bibfield  {journal} {\bibinfo  {journal} {Review of
  Scientific Instruments}\ }\textbf {\bibinfo {volume} {92}},\ \bibinfo {pages}
  {041101} (\bibinfo {year} {2021})},\ \Eprint
  {http://arxiv.org/abs/https://doi.org/10.1063/5.0023103}
  {https://doi.org/10.1063/5.0023103} \BibitemShut {NoStop}%
\bibitem [{\citenamefont {Yin}\ \emph {et~al.}(2017)\citenamefont {Yin},
  \citenamefont {Cao}, \citenamefont {Li}, \citenamefont {Liao}, \citenamefont
  {Zhang}, \citenamefont {Ren}, \citenamefont {Cai}, \citenamefont {Liu},
  \citenamefont {Li}, \citenamefont {Dai} \emph {et~al.}}]{yin2017satellite}%
  \BibitemOpen
  \bibfield  {author} {\bibinfo {author} {\bibfnamefont {J.}~\bibnamefont
  {Yin}}, \bibinfo {author} {\bibfnamefont {Y.}~\bibnamefont {Cao}}, \bibinfo
  {author} {\bibfnamefont {Y.-H.}\ \bibnamefont {Li}}, \bibinfo {author}
  {\bibfnamefont {S.-K.}\ \bibnamefont {Liao}}, \bibinfo {author}
  {\bibfnamefont {L.}~\bibnamefont {Zhang}}, \bibinfo {author} {\bibfnamefont
  {J.-G.}\ \bibnamefont {Ren}}, \bibinfo {author} {\bibfnamefont {W.-Q.}\
  \bibnamefont {Cai}}, \bibinfo {author} {\bibfnamefont {W.-Y.}\ \bibnamefont
  {Liu}}, \bibinfo {author} {\bibfnamefont {B.}~\bibnamefont {Li}}, \bibinfo
  {author} {\bibfnamefont {H.}~\bibnamefont {Dai}},  \emph {et~al.},\
  }\href@noop {} {\bibfield  {journal} {\bibinfo  {journal} {Science}\ }\textbf
  {\bibinfo {volume} {356}},\ \bibinfo {pages} {1140} (\bibinfo {year}
  {2017})}\BibitemShut {NoStop}%
\bibitem [{\citenamefont {Valivarthi}\ \emph {et~al.}(2016)\citenamefont
  {Valivarthi}, \citenamefont {Zhou}, \citenamefont {Aguilar}, \citenamefont
  {Verma}, \citenamefont {Marsili}, \citenamefont {Shaw}, \citenamefont {Nam},
  \citenamefont {Oblak}, \citenamefont {Tittel} \emph
  {et~al.}}]{valivarthi2016quantum}%
  \BibitemOpen
  \bibfield  {author} {\bibinfo {author} {\bibfnamefont {R.}~\bibnamefont
  {Valivarthi}}, \bibinfo {author} {\bibfnamefont {Q.}~\bibnamefont {Zhou}},
  \bibinfo {author} {\bibfnamefont {G.~H.}\ \bibnamefont {Aguilar}}, \bibinfo
  {author} {\bibfnamefont {V.~B.}\ \bibnamefont {Verma}}, \bibinfo {author}
  {\bibfnamefont {F.}~\bibnamefont {Marsili}}, \bibinfo {author} {\bibfnamefont
  {M.~D.}\ \bibnamefont {Shaw}}, \bibinfo {author} {\bibfnamefont {S.~W.}\
  \bibnamefont {Nam}}, \bibinfo {author} {\bibfnamefont {D.}~\bibnamefont
  {Oblak}}, \bibinfo {author} {\bibfnamefont {W.}~\bibnamefont {Tittel}},
  \emph {et~al.},\ }\href@noop {} {\bibfield  {journal} {\bibinfo  {journal}
  {Nature Photonics}\ }\textbf {\bibinfo {volume} {10}},\ \bibinfo {pages}
  {676} (\bibinfo {year} {2016})}\BibitemShut {NoStop}%
\bibitem [{\citenamefont {K\"onig}\ \emph {et~al.}(2005)\citenamefont
  {K\"onig}, \citenamefont {Mason}, \citenamefont {Wong},\ and\ \citenamefont
  {Albota}}]{PhysRevA.71.033805}%
  \BibitemOpen
  \bibfield  {author} {\bibinfo {author} {\bibfnamefont {F.}~\bibnamefont
  {K\"onig}}, \bibinfo {author} {\bibfnamefont {E.~J.}\ \bibnamefont {Mason}},
  \bibinfo {author} {\bibfnamefont {F.~N.~C.}\ \bibnamefont {Wong}}, \ and\
  \bibinfo {author} {\bibfnamefont {M.~A.}\ \bibnamefont {Albota}},\ }\href
  {\doibase 10.1103/PhysRevA.71.033805} {\bibfield  {journal} {\bibinfo
  {journal} {Phys. Rev. A}\ }\textbf {\bibinfo {volume} {71}},\ \bibinfo
  {pages} {033805} (\bibinfo {year} {2005})}\BibitemShut {NoStop}%
\bibitem [{\citenamefont {Clausen}\ \emph {et~al.}(2014)\citenamefont
  {Clausen}, \citenamefont {Bussi{\`{e}}res}, \citenamefont {Tiranov},
  \citenamefont {Herrmann}, \citenamefont {Silberhorn}, \citenamefont {Sohler},
  \citenamefont {Afzelius},\ and\ \citenamefont {Gisin}}]{Clausen_2014}%
  \BibitemOpen
  \bibfield  {author} {\bibinfo {author} {\bibfnamefont {C.}~\bibnamefont
  {Clausen}}, \bibinfo {author} {\bibfnamefont {F.}~\bibnamefont
  {Bussi{\`{e}}res}}, \bibinfo {author} {\bibfnamefont {A.}~\bibnamefont
  {Tiranov}}, \bibinfo {author} {\bibfnamefont {H.}~\bibnamefont {Herrmann}},
  \bibinfo {author} {\bibfnamefont {C.}~\bibnamefont {Silberhorn}}, \bibinfo
  {author} {\bibfnamefont {W.}~\bibnamefont {Sohler}}, \bibinfo {author}
  {\bibfnamefont {M.}~\bibnamefont {Afzelius}}, \ and\ \bibinfo {author}
  {\bibfnamefont {N.}~\bibnamefont {Gisin}},\ }\href {\doibase
  10.1088/1367-2630/16/9/093058} {\bibfield  {journal} {\bibinfo  {journal}
  {New Journal of Physics}\ }\textbf {\bibinfo {volume} {16}},\ \bibinfo
  {pages} {093058} (\bibinfo {year} {2014})}\BibitemShut {NoStop}%
\bibitem [{\citenamefont {Kim}\ \emph {et~al.}(2006)\citenamefont {Kim},
  \citenamefont {Fiorentino},\ and\ \citenamefont {Wong}}]{PhysRevA.73.012316}%
  \BibitemOpen
  \bibfield  {author} {\bibinfo {author} {\bibfnamefont {T.}~\bibnamefont
  {Kim}}, \bibinfo {author} {\bibfnamefont {M.}~\bibnamefont {Fiorentino}}, \
  and\ \bibinfo {author} {\bibfnamefont {F.~N.~C.}\ \bibnamefont {Wong}},\
  }\href {\doibase 10.1103/PhysRevA.73.012316} {\bibfield  {journal} {\bibinfo
  {journal} {Phys. Rev. A}\ }\textbf {\bibinfo {volume} {73}},\ \bibinfo
  {pages} {012316} (\bibinfo {year} {2006})}\BibitemShut {NoStop}%
\bibitem [{\citenamefont {Fedrizzi}\ \emph {et~al.}(2007)\citenamefont
  {Fedrizzi}, \citenamefont {Herbst}, \citenamefont {Poppe}, \citenamefont
  {Jennewein},\ and\ \citenamefont {Zeilinger}}]{fedrizzi2007wavelength}%
  \BibitemOpen
  \bibfield  {author} {\bibinfo {author} {\bibfnamefont {A.}~\bibnamefont
  {Fedrizzi}}, \bibinfo {author} {\bibfnamefont {T.}~\bibnamefont {Herbst}},
  \bibinfo {author} {\bibfnamefont {A.}~\bibnamefont {Poppe}}, \bibinfo
  {author} {\bibfnamefont {T.}~\bibnamefont {Jennewein}}, \ and\ \bibinfo
  {author} {\bibfnamefont {A.}~\bibnamefont {Zeilinger}},\ }\href@noop {}
  {\bibfield  {journal} {\bibinfo  {journal} {Optics Express}\ }\textbf
  {\bibinfo {volume} {15}},\ \bibinfo {pages} {15377} (\bibinfo {year}
  {2007})}\BibitemShut {NoStop}%
\bibitem [{\citenamefont {de~Chatellus}\ \emph {et~al.}(2006)\citenamefont
  {de~Chatellus}, \citenamefont {Sergienko}, \citenamefont {Saleh},
  \citenamefont {Teich},\ and\ \citenamefont {Giuseppe}}]{deChatellus:06}%
  \BibitemOpen
  \bibfield  {author} {\bibinfo {author} {\bibfnamefont {H.~G.}\ \bibnamefont
  {de~Chatellus}}, \bibinfo {author} {\bibfnamefont {A.~V.}\ \bibnamefont
  {Sergienko}}, \bibinfo {author} {\bibfnamefont {B.~E.~A.}\ \bibnamefont
  {Saleh}}, \bibinfo {author} {\bibfnamefont {M.~C.}\ \bibnamefont {Teich}}, \
  and\ \bibinfo {author} {\bibfnamefont {G.~D.}\ \bibnamefont {Giuseppe}},\
  }\href {\doibase 10.1364/OE.14.010060} {\bibfield  {journal} {\bibinfo
  {journal} {Opt. Express}\ }\textbf {\bibinfo {volume} {14}},\ \bibinfo
  {pages} {10060} (\bibinfo {year} {2006})}\BibitemShut {NoStop}%
\bibitem [{\citenamefont {Pelton}\ \emph {et~al.}(2004)\citenamefont {Pelton},
  \citenamefont {Marsden}, \citenamefont {Ljunggren}, \citenamefont {Tengner},
  \citenamefont {Karlsson}, \citenamefont {Fragemann}, \citenamefont
  {Canalias},\ and\ \citenamefont {Laurell}}]{Pelton:04}%
  \BibitemOpen
  \bibfield  {author} {\bibinfo {author} {\bibfnamefont {M.}~\bibnamefont
  {Pelton}}, \bibinfo {author} {\bibfnamefont {P.}~\bibnamefont {Marsden}},
  \bibinfo {author} {\bibfnamefont {D.}~\bibnamefont {Ljunggren}}, \bibinfo
  {author} {\bibfnamefont {M.}~\bibnamefont {Tengner}}, \bibinfo {author}
  {\bibfnamefont {A.}~\bibnamefont {Karlsson}}, \bibinfo {author}
  {\bibfnamefont {A.}~\bibnamefont {Fragemann}}, \bibinfo {author}
  {\bibfnamefont {C.}~\bibnamefont {Canalias}}, \ and\ \bibinfo {author}
  {\bibfnamefont {F.}~\bibnamefont {Laurell}},\ }\href {\doibase
  10.1364/OPEX.12.003573} {\bibfield  {journal} {\bibinfo  {journal} {Opt.
  Express}\ }\textbf {\bibinfo {volume} {12}},\ \bibinfo {pages} {3573}
  (\bibinfo {year} {2004})}\BibitemShut {NoStop}%
\bibitem [{\citenamefont {H\"{u}bel}\ \emph {et~al.}(2007)\citenamefont
  {H\"{u}bel}, \citenamefont {Vanner}, \citenamefont {Lederer}, \citenamefont
  {Blauensteiner}, \citenamefont {Lor\"{u}nser}, \citenamefont {Poppe},\ and\
  \citenamefont {Zeilinger}}]{Hubel:07}%
  \BibitemOpen
  \bibfield  {author} {\bibinfo {author} {\bibfnamefont {H.}~\bibnamefont
  {H\"{u}bel}}, \bibinfo {author} {\bibfnamefont {M.~R.}\ \bibnamefont
  {Vanner}}, \bibinfo {author} {\bibfnamefont {T.}~\bibnamefont {Lederer}},
  \bibinfo {author} {\bibfnamefont {B.}~\bibnamefont {Blauensteiner}}, \bibinfo
  {author} {\bibfnamefont {T.}~\bibnamefont {Lor\"{u}nser}}, \bibinfo {author}
  {\bibfnamefont {A.}~\bibnamefont {Poppe}}, \ and\ \bibinfo {author}
  {\bibfnamefont {A.}~\bibnamefont {Zeilinger}},\ }\href {\doibase
  10.1364/OE.15.007853} {\bibfield  {journal} {\bibinfo  {journal} {Opt.
  Express}\ }\textbf {\bibinfo {volume} {15}},\ \bibinfo {pages} {7853}
  (\bibinfo {year} {2007})}\BibitemShut {NoStop}%
\bibitem [{\citenamefont {Fiorentino}\ and\ \citenamefont
  {Beausoleil}(2008)}]{Fiorentino:08}%
  \BibitemOpen
  \bibfield  {author} {\bibinfo {author} {\bibfnamefont {M.}~\bibnamefont
  {Fiorentino}}\ and\ \bibinfo {author} {\bibfnamefont {R.~G.}\ \bibnamefont
  {Beausoleil}},\ }\href {\doibase 10.1364/OE.16.020149} {\bibfield  {journal}
  {\bibinfo  {journal} {Opt. Express}\ }\textbf {\bibinfo {volume} {16}},\
  \bibinfo {pages} {20149} (\bibinfo {year} {2008})}\BibitemShut {NoStop}%
\bibitem [{\citenamefont {Sauge}\ \emph {et~al.}(2008)\citenamefont {Sauge},
  \citenamefont {Swillo}, \citenamefont {Tengner},\ and\ \citenamefont
  {Karlsson}}]{Sauge:08}%
  \BibitemOpen
  \bibfield  {author} {\bibinfo {author} {\bibfnamefont {S.}~\bibnamefont
  {Sauge}}, \bibinfo {author} {\bibfnamefont {M.}~\bibnamefont {Swillo}},
  \bibinfo {author} {\bibfnamefont {M.}~\bibnamefont {Tengner}}, \ and\
  \bibinfo {author} {\bibfnamefont {A.}~\bibnamefont {Karlsson}},\ }\href
  {\doibase 10.1364/OE.16.009701} {\bibfield  {journal} {\bibinfo  {journal}
  {Opt. Express}\ }\textbf {\bibinfo {volume} {16}},\ \bibinfo {pages} {9701}
  (\bibinfo {year} {2008})}\BibitemShut {NoStop}%
\bibitem [{\citenamefont {Lee}\ \emph {et~al.}(2021)\citenamefont {Lee},
  \citenamefont {Xie}, \citenamefont {Tannous},\ and\ \citenamefont
  {Jennewein}}]{Lee_2021}%
  \BibitemOpen
  \bibfield  {author} {\bibinfo {author} {\bibfnamefont {Y.~S.}\ \bibnamefont
  {Lee}}, \bibinfo {author} {\bibfnamefont {M.}~\bibnamefont {Xie}}, \bibinfo
  {author} {\bibfnamefont {R.}~\bibnamefont {Tannous}}, \ and\ \bibinfo
  {author} {\bibfnamefont {T.}~\bibnamefont {Jennewein}},\ }\href {\doibase
  10.1088/2058-9565/abd151} {\bibfield  {journal} {\bibinfo  {journal} {Quantum
  Science and Technology}\ }\textbf {\bibinfo {volume} {6}},\ \bibinfo {pages}
  {025004} (\bibinfo {year} {2021})}\BibitemShut {NoStop}%
\bibitem [{\citenamefont {Hentschel}\ \emph {et~al.}(2009)\citenamefont
  {Hentschel}, \citenamefont {H\"{u}bel}, \citenamefont {Poppe},\ and\
  \citenamefont {Zeilinger}}]{Hentschel:09}%
  \BibitemOpen
  \bibfield  {author} {\bibinfo {author} {\bibfnamefont {M.}~\bibnamefont
  {Hentschel}}, \bibinfo {author} {\bibfnamefont {H.}~\bibnamefont
  {H\"{u}bel}}, \bibinfo {author} {\bibfnamefont {A.}~\bibnamefont {Poppe}}, \
  and\ \bibinfo {author} {\bibfnamefont {A.}~\bibnamefont {Zeilinger}},\ }\href
  {\doibase 10.1364/OE.17.023153} {\bibfield  {journal} {\bibinfo  {journal}
  {Opt. Express}\ }\textbf {\bibinfo {volume} {17}},\ \bibinfo {pages} {23153}
  (\bibinfo {year} {2009})}\BibitemShut {NoStop}%
\bibitem [{\citenamefont {Steinlechner}\ \emph {et~al.}(2012)\citenamefont
  {Steinlechner}, \citenamefont {Trojek}, \citenamefont {Jofre}, \citenamefont
  {Weier}, \citenamefont {Perez}, \citenamefont {Jennewein}, \citenamefont
  {Ursin}, \citenamefont {Rarity}, \citenamefont {Mitchell}, \citenamefont
  {Torres} \emph {et~al.}}]{steinlechner2012high}%
  \BibitemOpen
  \bibfield  {author} {\bibinfo {author} {\bibfnamefont {F.}~\bibnamefont
  {Steinlechner}}, \bibinfo {author} {\bibfnamefont {P.}~\bibnamefont
  {Trojek}}, \bibinfo {author} {\bibfnamefont {M.}~\bibnamefont {Jofre}},
  \bibinfo {author} {\bibfnamefont {H.}~\bibnamefont {Weier}}, \bibinfo
  {author} {\bibfnamefont {D.}~\bibnamefont {Perez}}, \bibinfo {author}
  {\bibfnamefont {T.}~\bibnamefont {Jennewein}}, \bibinfo {author}
  {\bibfnamefont {R.}~\bibnamefont {Ursin}}, \bibinfo {author} {\bibfnamefont
  {J.}~\bibnamefont {Rarity}}, \bibinfo {author} {\bibfnamefont {M.~W.}\
  \bibnamefont {Mitchell}}, \bibinfo {author} {\bibfnamefont {J.~P.}\
  \bibnamefont {Torres}},  \emph {et~al.},\ }\href@noop {} {\bibfield
  {journal} {\bibinfo  {journal} {Optics express}\ }\textbf {\bibinfo {volume}
  {20}},\ \bibinfo {pages} {9640} (\bibinfo {year} {2012})}\BibitemShut
  {NoStop}%
\bibitem [{\citenamefont {Schneeloch}\ \emph {et~al.}(2019)\citenamefont
  {Schneeloch}, \citenamefont {Knarr}, \citenamefont {Bogorin}, \citenamefont
  {Levangie}, \citenamefont {Tison}, \citenamefont {Frank}, \citenamefont
  {Howland}, \citenamefont {Fanto},\ and\ \citenamefont
  {Alsing}}]{schneeloch2019introduction}%
  \BibitemOpen
  \bibfield  {author} {\bibinfo {author} {\bibfnamefont {J.}~\bibnamefont
  {Schneeloch}}, \bibinfo {author} {\bibfnamefont {S.~H.}\ \bibnamefont
  {Knarr}}, \bibinfo {author} {\bibfnamefont {D.~F.}\ \bibnamefont {Bogorin}},
  \bibinfo {author} {\bibfnamefont {M.~L.}\ \bibnamefont {Levangie}}, \bibinfo
  {author} {\bibfnamefont {C.~C.}\ \bibnamefont {Tison}}, \bibinfo {author}
  {\bibfnamefont {R.}~\bibnamefont {Frank}}, \bibinfo {author} {\bibfnamefont
  {G.~A.}\ \bibnamefont {Howland}}, \bibinfo {author} {\bibfnamefont {M.~L.}\
  \bibnamefont {Fanto}}, \ and\ \bibinfo {author} {\bibfnamefont {P.~M.}\
  \bibnamefont {Alsing}},\ }\href@noop {} {\bibfield  {journal} {\bibinfo
  {journal} {Journal of Optics}\ }\textbf {\bibinfo {volume} {21}},\ \bibinfo
  {pages} {043501} (\bibinfo {year} {2019})}\BibitemShut {NoStop}%
\bibitem [{\citenamefont {Clauser}\ \emph {et~al.}(1969)\citenamefont
  {Clauser}, \citenamefont {Horne}, \citenamefont {Shimony},\ and\
  \citenamefont {Holt}}]{clauser1969proposed}%
  \BibitemOpen
  \bibfield  {author} {\bibinfo {author} {\bibfnamefont {J.~F.}\ \bibnamefont
  {Clauser}}, \bibinfo {author} {\bibfnamefont {M.~A.}\ \bibnamefont {Horne}},
  \bibinfo {author} {\bibfnamefont {A.}~\bibnamefont {Shimony}}, \ and\
  \bibinfo {author} {\bibfnamefont {R.~A.}\ \bibnamefont {Holt}},\ }\href@noop
  {} {\bibfield  {journal} {\bibinfo  {journal} {Physical review letters}\
  }\textbf {\bibinfo {volume} {23}},\ \bibinfo {pages} {880} (\bibinfo {year}
  {1969})}\BibitemShut {NoStop}%
\bibitem [{\citenamefont {James}\ \emph {et~al.}(2001)\citenamefont {James},
  \citenamefont {Kwiat}, \citenamefont {Munro},\ and\ \citenamefont
  {White}}]{james2001measurement}%
  \BibitemOpen
  \bibfield  {author} {\bibinfo {author} {\bibfnamefont {D.~F.}\ \bibnamefont
  {James}}, \bibinfo {author} {\bibfnamefont {P.~G.}\ \bibnamefont {Kwiat}},
  \bibinfo {author} {\bibfnamefont {W.~J.}\ \bibnamefont {Munro}}, \ and\
  \bibinfo {author} {\bibfnamefont {A.~G.}\ \bibnamefont {White}},\ }\href@noop
  {} {\bibfield  {journal} {\bibinfo  {journal} {Physical Review A}\ }\textbf
  {\bibinfo {volume} {64}},\ \bibinfo {pages} {052312} (\bibinfo {year}
  {2001})}\BibitemShut {NoStop}%
\bibitem [{\citenamefont {Horn}\ and\ \citenamefont
  {Jennewein}(2019)}]{Horn:19}%
  \BibitemOpen
  \bibfield  {author} {\bibinfo {author} {\bibfnamefont {R.}~\bibnamefont
  {Horn}}\ and\ \bibinfo {author} {\bibfnamefont {T.}~\bibnamefont
  {Jennewein}},\ }\href {\doibase 10.1364/OE.27.017369} {\bibfield  {journal}
  {\bibinfo  {journal} {Opt. Express}\ }\textbf {\bibinfo {volume} {27}},\
  \bibinfo {pages} {17369} (\bibinfo {year} {2019})}\BibitemShut {NoStop}%
\end{thebibliography}%

\end{document}